\documentclass[%
 reprint,
 amsmath,amssymb,
 aps,
]{revtex4-1}
\usepackage{graphicx}
\usepackage{dcolumn}
\usepackage{bm}

\begin{document}

\title{Identification of the crystal structures of two superconducting phases for potassium-doped picene}

\author{Xun-Wang Yan$^{1,2,3,4}$}
\author{Zhongbing Huang$^{1,2}$}
\email{huangzb@hubu.edu.cn}
\author{Hai-Qing Lin$^1$}
\affiliation{$^1$Beijing Computational Science Research Center, Beijing 100084, China}
\affiliation{$^2$Faculty of Physics and Electronic Technology, Hubei University, Wuhan 430062, China}
\affiliation{$^3$State Key Laboratory of Theoretical Physics, Institute of Theoretical Physics,
Chinese Academy of Science, Beijing  100190, China }
\affiliation{$^4$School of physics and electrical engineering, Anyang Normal University, Henan 455000, China}
\date{\today}

\begin{abstract}

By the first principles calculations and X-ray diffraction simulations, we perform a systematic
study of multiple superconducting phases of potassium-doped picene [Nature {\bf464} 76, 2010].
The combination of optimized lattice parameters, formation energy, and simulated X-ray diffraction
spectra indicates that the superconducting phase with transition temperature (T$_c$) of 7 K
corresponds to a charge doped K$_2$picene, which is semiconducting with a 0.1 eV energy gap, while
the superconducting phase with T$_c$=18 K corresponds to a metallic K$_3$picene. The distinct
crystal and electronic structures of K$_2$picene and K$_3$picene provide a reasonable explanation
of two superconducting phases in potassium-doped picene.

\end{abstract}

\maketitle

Superconductivity in organic materials are among the most fascinating phenomena in condensed matter physics.
The recent discovery of superconductivity in potassium-doped picene (C$_{22}$H$_{11}$) \cite{Mitsuhashi2010}
stimulated great interest of researchers to investigate superconductivity in aromatic solid materials.
A few kinds of aromatic molecules, including picene (C$_{22}$H$_{11}$), phenanthrene (C$_{14}$H$_{10}$)
\cite{Wang2011}, coronene (C$_{24}$H$_{12}$) and dibenzopentacene (C$_{30}$H$_{18}$)\cite{Xue2012}
have been used to synthesize aromatic superconductors by intercalating metal atoms into the molecular
crystals \cite{Wang2011a,Wang2012}. Experimental measurements \cite{Kambe2012,Caputo2012, Ruff2013,Teranishi2013}
and theoretical calculations \cite{Kosugi2011,DeAndres2011a,DeAndres2011,PhysRevB.88.115106,Yan2013} showed that
there exist electronic correlations, magnetism \cite{Kim2011,Giovannetti2011,Kim2013,Huang2012,Verges2012},
and pronounced electron-phonon coupling \cite{Kato2011} in aromatic superconductors, which are very similar to
the situations of high-T$_c$ cuprates and iron-based superconductors. Hence, it is challenging to understand
the superconducting (SC) mechanism in aromatic superconductors.

One of the most intriguing properties of aromatic superconductors is the existence of multiple SC phases in
K-doped picene, coronene, and dibenzopentacene. In particular, two different SC phases in K-doped picene,
one with T$_c$ $\sim$ 7 K and the other as high as 18 K, have attracted considerable attention.
Currently, the crystal structures of K-doped picene remain unclear. On the experimental side, the measured
lattice parameters showed large discrepancies for the samples prepared by solid-state reaction method and
solution method (see Table~I), and it is difficult to derive the positions of intercalated K atoms. On the
theoretical side, the optimized crystal structures from the first principles calculations deviate dramatically
from the experimental ones: the experimental lattice parameter a is much larger than b, while a and b take
similar values from the first principles simulations.

Since the determination of atomic structure of materials is a prerequisite to explore their electronic,
magnetic, and SC properties, it is urgent to identify the crystal structures of aromatic superconductors,
especially for two SC phases in K-doped picene, where various experimental data have been accumulated.
Besides aromatic superconductors, the identification of crystal structure is crucial for understanding
the SC property of other alkali metal intercalated layered materials, including iron-based superconductors,
graphite, and graphene. Interesting, in FeSe, alkali metal (Li, Na, or K) has been intercalated
into the space between FeSe layers to optimize the SC property, which also leads to multiple SC phases
at T$_c$=30$\sim$46K.

In this Letter, we have performed a systematic study of two SC phases in K-doped picene by combining the
first principles calculations and X-ray diffraction (XRD) simulations. Based on the optimized lattice
parameters, formation energy, and XRD spectra, we find that the SC phase with T$_c$ $\sim$ 7 K is indeed
related to a semiconducting K$_2$picene, whereas the SC phase with T$_c$ = 18 K corresponds to a metallic
K$_3$picene. This is the first time to identify theoretically the difference of crystal structures and dopant
concentrations between two different SC phases. The relatively large density of states of 10.4 states/eV
at the Fermi level in K$_3$picene provide a reasonable explanation of higher T$_c$ in the T$_c$ = 18 K SC phase.
In addition, we point out that an inclusion of KOH in SC samples not only explains the experimental XRD peaks,
but also demonstrates that preventing KOH generation in experiment is an important measure to improve the
sample quality of K-doped aromatic superconductors.

In our calculations the plane wave basis and pseudopotential method was used. The psudopotentials
are supplied by Vienna Ab initio simulation package (VASP) website \cite{PhysRevB.47.558, PhysRevB.54.11169},
which adopt the projector augmented-wave method (PAW) \cite{PhysRevB.50.17953} and Perdew-Burke-Ernzerhof (PBE)
type generalized gradient approximation (GGA) \cite{PhysRevLett.77.3865}. The plane wave basis cutoff is set
to 300 eV. The Gaussian broadening technique was used and a mesh of $4\times 6\times 4$ k-points were sampled
for the Brillouin-zone integration. The convergence thresholds of the total energy, force on atom and pressure
on cell are 10$^{-4}$ eV, 0.01 eV/\AA and 0.1 KBar, respectively. The XRD sepectra were simulated by Mercury
program \cite{mercury}.

\begin{figure}.
\includegraphics[width=7.0cm]{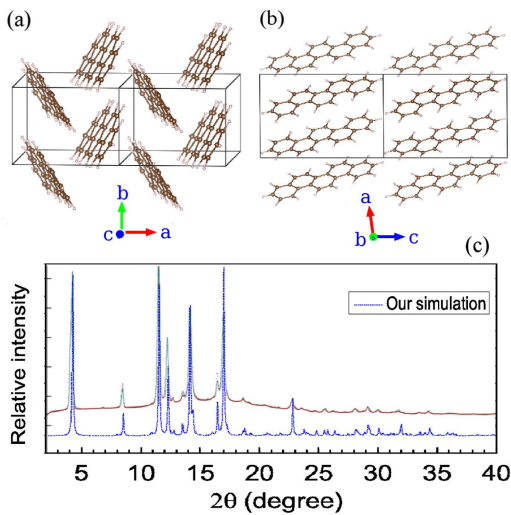}
\caption{(Color online)
The intralayer arrangement of molecules (a) shown in a 2 $\times$ 1 $\times$ 1 supercell, and the interlayer
arrangement of molecules (b) shown in a 1 $\times$ 1 $\times$ 2 supercell. (c) our simulated XRD spectra
(blue line), and the experimental one (black circles and red line) adapted by permission from Macmillan
Publishers Ltd: Nature {\bf 464} 7285 (Fig. S7), R. Mitsuhashi {\it et al}, copyright (2010).
The used X-ray wave length is 0.99937 \AA.} \label{pristine-struct}
\end{figure}

We first study the structure of pristine picene, which is the basis for the structure exploration of K-doped counterpart.
Picene molecule is a flat hydrocarbon composed of five fused benzene rings. The molecular crystal of pristine picene
crystallizes in the space group $P2_{1}$ and each unit cell contains two molecules, which are arranged in a herringbone
pattern to form a layer parallel to the $ab$ plane, and then molecular layers are stacked along the $c$ axis, as shown
in Figs. \ref{pristine-struct}(a) and \ref{pristine-struct}(b). The optimized lattice parameters $a, b, c, \beta$ are
8.489 \AA, 6.124 \AA, 13.444 \AA, 90.43$^{\circ}$, in excellently agreement with the experimental values 8.472 \AA,
6.17\AA, 13.538 \AA, 90.84$^{\circ}$). The simulated XRD spectra of picene shown in Fig. \ref{pristine-struct}(c) are
in perfectly consistent with the measured ones. These results provide a fundamental guarantee for the applicability
of C and H pseudopotentials, the reasonability of k-point mesh, and plane wave basis cutoff in our following
calculations of K-doped picene.

\begin{figure}.
\includegraphics[width=7.0cm]{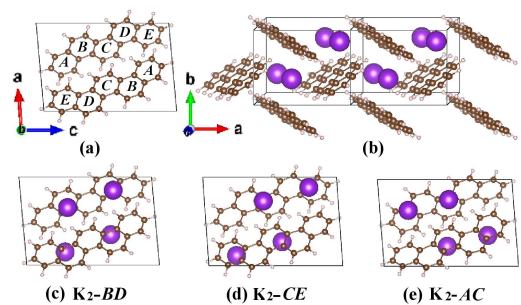}
\caption{(Color online)
(a) Five favorable positions of K atom in the interstitial space viewed along $b$ axis, named as $A, B, C, D, E$
corresponding to five benzene rings repectively. (b) A schematic diagram of K atom arrangement for K-doped picene.
(c), (d) and (e) are three possible structure phases for K$_2$picene.} \label{K2ABC}.
\end{figure}

\begin{figure}.
\includegraphics[width=7.0cm]{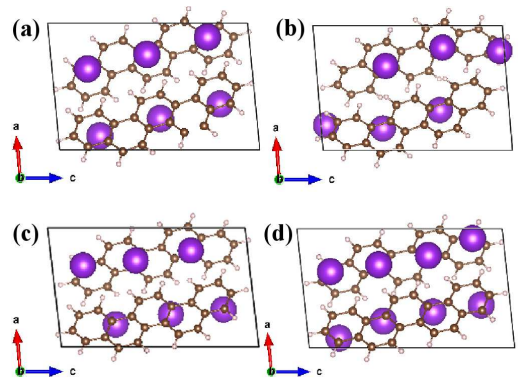}
\caption{(Color online)
 A schematic diagram of three possible structure phases for K$_3$picene, named as K$_3$-\uppercase\expandafter{\romannumeral1} (a), K$_3$-\uppercase\expandafter{\romannumeral2} (b), K$_3$-\uppercase\expandafter{\romannumeral3} (c), and  K$_4$picene (d).}
\label{K3-123}
\end{figure}

Experimental results indicated that K-doped picene also crystallizes in the space group $P2_{1}$ and two molecules in
a unit cell arrange in the herringbone pattern. In order to distinguish the K atom positions in the molecular crystal,
we define the interstitial space enclosed by four molecules in a molecular layer as a hole, and name the positions
corresponding to five benzene rings of a molecule as $A, B, C, D, E$, as shown in Fig. \ref{K2ABC}(a). When two K
atoms are intercalated into each hole, three kinds of possible configurations with $P2_{1}$ symmetry can be formed,
which are called as K$_2$-$BD$, K$_2$-$AC$, and K$_2$-$CE$ shown in Figs. \ref{K2ABC}(c), (d), and (e).
The configurations are named after their K atom positions. Fig. \ref{K2ABC}(b) displays the positions of K atoms and
molecules viewed along the $c$ axis for K$_2$-$BD$, and the other two phases have a similar arrangement.

The structural phases of K$_3$picene can be constructed by adding two K atoms to the unit cell of K$_2$picene phase.
On the basis of K$_2$-$AC$ or K$_2$-$CE$, the occupation with K atom at the $E$ or $A$ position lead to the
K$_3$-\uppercase\expandafter{\romannumeral1} phase, as shown in Fig. \ref{K3-123}(a). When two K atoms are added to
the K$_2$-$BD$ cell, there are two kinds of arrangement, $i.e.$, two K atoms are put outside $E$ and outside $A$,
corresponding to the initial configuration of phases K$_3$-\uppercase\expandafter{\romannumeral2} and
K$_3$-\uppercase\expandafter{\romannumeral3}. The structure of K$_3$-\uppercase\expandafter{\romannumeral2} is similar
to K$_2$K$_1$ in Ref. \cite{Kosugi2011} and two K atoms still sit outside $E$ positions after relaxation shown in
Fig. \ref{K3-123}(b). However, K atoms outside $A$ position in the K$_3$-\uppercase\expandafter{\romannumeral3} phase
have a large change and move into the hole after structure optimization shown in Fig. \ref{K3-123}(c).

\begin{table}
\caption{\label{K2ABC-latt} The optimized lattice parameters $a, b , c, \beta$, the fraction coordinations of
the doped K atoms and the space group of unit cell for K$_2$picene and K$_3$picene with different structural phases.
}
\begin{tabular}{l c c c c c}
\hline
\hline
         &a (\AA) &b (\AA) &c (\AA) &  $\beta$ ($^{\circ}$) & space group \\
\hline
K$_2$-$BD$          &8.766 &6.818 &13.166   &95.13  &P2$_1$\\
                &(0.3461 &     0.2917  &    0.6540) & \\
                &(0.1500  &    0.3017  &    0.2981) & \\
K$_2$-$CE$          &8.752&6.556 &13.293 &92.98  & P2$_1$\\
        &(0.2571  &    0.3062  &    0.5452 )& \\
        &(0.1073   &   0.2856  &    0.1980 )& \\
K$_2$-$AC$      & 8.651 &6.524  &13.306& 92.60  & P2$_1$ \\
        &(0.3150 &  0.3269  &   0.8168   )& \\
        &(0.2352 & 0.3291  &   0.5126  )& \\
Exp. \cite{Mitsuhashi2010}   &{\bf 8.707} &{\bf 5.912} & {\bf 12.97}   & {\bf 92.77}  & P2$_1$\\
K$_3$-I     & 8.675 &6.770  &13.669& 95.53  & P2$_1$ \\
          & (0.3611    &  0.3057 &     0.8222)  &\\
         & (0.2470    &  0.2913  &    0.5183)  & \\
         & (0.1098    &  0.2912   &   0.2074)  &\\
K$_3$-II     & 8.914 &6.793  &13.534& 94.72  & P2$_1$ \\
   & (0.3227  &    0.2953  &    0.6336)  &\\
   & (0.1753  &    0.2949  &    0.3250)  &\\
   & (0.2013   &   0.2372  &    0.0257)  &\\
K$_3$-III     & 8.523 &6.838  &14.058& 96.67  & P2$_1$ \\
          &  (0.2732    &  0.3048   &   0.5729) & \\
          &  (0.1904      & 0.2976    &  0.2882) &   \\
          &   (0.3153    &  0.2897   &   0.8434) &  \\
Exp. \cite{Kambe2012}   &{\bf 8.571} & {\bf 6.270} & {\bf 14.01}   & {\bf 91.68}  & P2$_1$\\
\hline
\hline
\end{tabular}
\end{table}

The optimized lattice parameters and fraction coordinations of K atoms are listed in Table.~\ref{K2ABC-latt}.
For K$_2$ picene, the energy of K$_2$-$BD$ is lower than that of K$_2$-$CE$ by 0.28 eV/molecule and that of
K$_2$-$AC$ by 0.41 eV/molecule, suggesting that the K$_2$-$BD$ phase is the most stable configuration for
K$_2$picene from the energetic point of view. As a result, the K$_2$-$BD$ phase has a higher probability of
existence in the experimental samples. Among the three phases for K$_3$picene, K$_3$-\uppercase\expandafter{\romannumeral1}
has the lowest energy, which is 0.46 eV and 0.022eV lower than K$_3$-\uppercase\expandafter{\romannumeral2}
and K$_3$-\uppercase\expandafter{\romannumeral3} per unit cell, respectively. Notice that K$_2$-$BD$ and
K$_3$-\uppercase\expandafter{\romannumeral1} are in good agreement with the K$_{2.9}$ sample~\cite{Mitsuhashi2010}
and the T$_c$=18 K sample~\cite{Kambe2012}, except for a bit longer b axis.

We also calculate the formation energy in term of the formula $E_{formation} = E_{Kx}-E_{pristine}-E_{dopant}$,
where $x$ stands for the number of K atoms and $E_{dopant}$ is the product of single atom energy in bulk metal
and atom number in a unit cell. For the K$_2$-$BD$ phase, the formation energy is -0.330 eV per K atom, which
suggests that it is easy to synthesize in experiment. The formation energy of the
K$_3$-\uppercase\expandafter{\romannumeral1} phase is -0.295 eV per K atom, comparable to the one for K$_2$picene,
indicating that K$_3$picene is another reasonable structure phase for K-doped picene. For K$_4$picene as shown
in Fig. \ref{K3-123}(d), the corresponding formation energy is -0.067 eV per K atom, much less than the ones
for K$_2$picene and K$_3$picene, indicating that K$_4$picene is not a stable structural phase.

\begin{figure}.
\includegraphics[width=7.0cm]{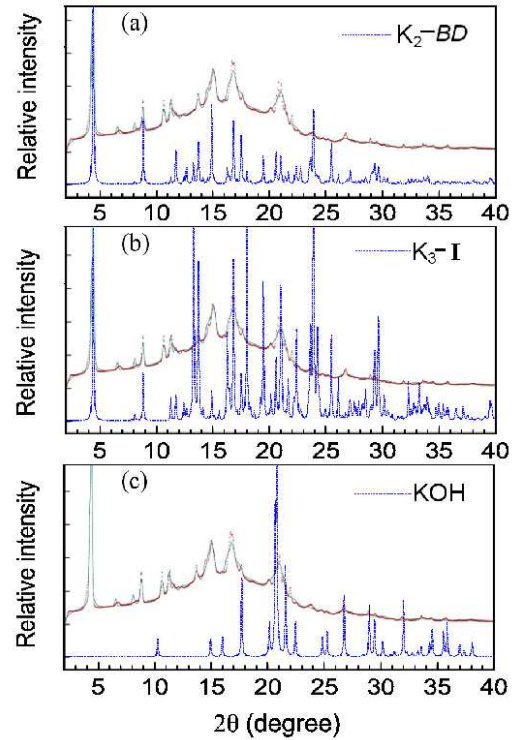}
\caption{(Color online)
The simulated XRD spectra of K$_2$-$BD$ phase (a) and K$_3$-\uppercase\expandafter{\romannumeral1} phase
(b) based on the experimental lattice parameters \cite{Mitsuhashi2010}. (c) The XRD spectra of KOH with the
experimental crystal structure. The experimental XRD spectra (black circles and red line) of K$_{2.9}$picene
adapted by permission from Macmillan Publishers Ltd: Nature {\bf 464} 7285 (Fig. S7), R. Mitsuhashi {\it et al},
copyright (2010). The used X-ray wave length is 0.99937 \AA.}  \label{K2-XRD}
\end{figure}

Having identifying the stable structure phases for K-doped picene, we now turn to identify the crystal structure
of experimental SC samples. In Fig.~\ref{K2-XRD}, we compare the simulated XRD spectra of K$_2$-$BD$ and
K$_3$-\uppercase\expandafter{\romannumeral1} with the measured results for the sample with T$_c$=7K~\cite{Mitsuhashi2010}.
For both K$_2$-$BD$ and K$_3$-\uppercase\expandafter{\romannumeral1}, we fix the lattice parameters at experimental
values to relax the inner atoms. One can clearly see that the simulated XRD spectra of K$_2$-$BD$ is in good agreement
with the measured one, which can be seen from the positions and strength of the main XRD peaks shown in Fig. \ref{K2-XRD}(a).
The $z$ fraction coordinations for two independent K atoms are (0.3127, 0.6626), which is close to ($\frac{1}{3}$,
$\frac{2}{3}$), similar to the $z$ coordinations (0.2981, 0.6540) in the full relaxation case. In contrast,
the XRD structure of K$_3$-\uppercase\expandafter{\romannumeral1} is not consistent with the expeimental one.
Based on the above analysis on energy and XRD spectra, we can draw a conclusion: the T$_c$ $\sim$ 7 K SC phase
corresponds to K$_2$picene with a K$_2$-$BD$ structure.

\begin{figure}.
\includegraphics[width=7.0cm]{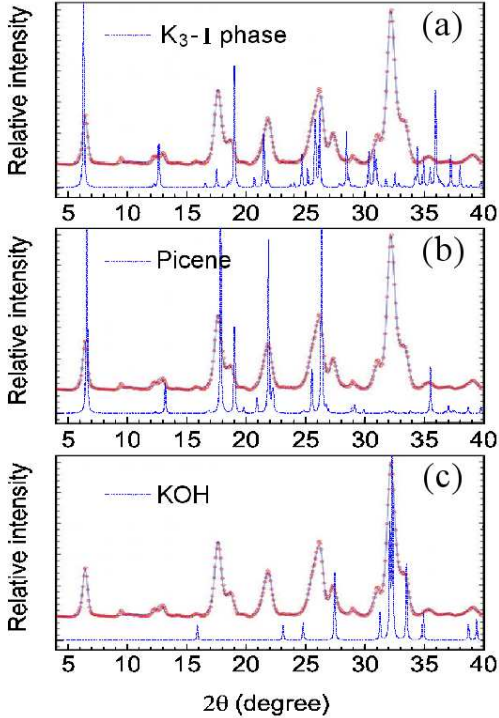}
\caption{(Color online)
The simulated XRD spectra of K$_3$-\uppercase\expandafter{\romannumeral1} phase (a) based on the experimental
lattice parameters \cite{Kambe2012} and the XRD spectra of pristine picene (b) and KOH (c) with the experimental
crystal structure, ploted with bright blue line. The experimental XRD spectra (red circles and blue line) of K$_{3}$
picene is also presented in three panels, which data is from Ref. \cite{Kambe2012}. The used X-ray wave length is 1.54056 \AA.}
\label{K3-XRD}
\end{figure}

In Fig.\ref{K3-XRD}, the simulated XRD spectra are compared with the experimental measurement for the T$_c$ = 18 K phase
by a solution method \cite{Kambe2012}. The XRD spectra of K$_3$-\uppercase\expandafter{\romannumeral1} based on the experimental
lattice parameters as displayed in Fig.\ref{K3-XRD}(a) show that the peak positions below 28 degree are in accordance with
the experiment. Our results (not shown here) also indicate that the XRD spectra for K$_3$-\uppercase\expandafter{\romannumeral2}
and K$_3$-\uppercase\expandafter{\romannumeral3} have a stronger discrepancy to the experiment than
K$_3$-\uppercase\expandafter{\romannumeral1}. For this reason, the K$_3$-\uppercase\expandafter{\romannumeral1} phase is the
most possible configuration corresponding to T$_c$ = 18 K phase. As seen from Figs.\ref{K3-XRD}(b) and (c), the XRD peaks of
pristine picene and KOH have a good agreement with the experimental ones, suggesting that pristine picene and KOH have a
considerable portion in the SC sample. This can explain the reason of XRD spectra difference in Fig. \ref{K3-XRD}(a) and
the fact of small superconductive shielding fraction.

\begin{figure}
  \includegraphics[width=7.0cm]{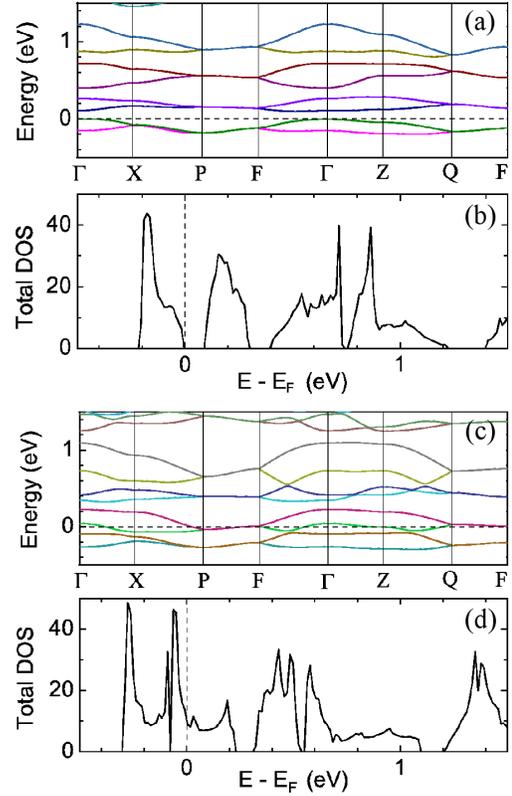}
  \caption{Energy bands and density of states for K$_2$-$BD$ phase (a) (b) and K$_3$-\uppercase\expandafter{\romannumeral1}
  (c) (d). Fermi energy is set to zero.} \label{K2-band-dos}
\end{figure}

The band structure and DOS of K$_2$-$BD$ and K$_3$-\uppercase\expandafter{\romannumeral1} are presented in Fig. \ref{K2-band-dos}.
On the  whole the bands are flat and little dispersive. For K$_2$-$BD$ phase, two bands below Fermi level are mainly from the
lowest unoccupied molecular orbitals (LUMO) of two molecules in a unit cell in Fig. \ref{K2-band-dos} (a), which are occupied
as a result of the intercalation of two K atoms per molecule. For K$_3$-\uppercase\expandafter{\romannumeral1} phase of
K$_3$picene, the Fermi level crosses the two bands coming from LUMO+1 orbitals.
K$_2$-$BD$ phase is a semiconductor with an energy gap of 0.1 eV, which is the parent compounds for T$_c$ = 7 K SC phase
in terms of above analysis. When the concentration of K atom has few percents deviation from 2 in K$_2$-$BD$ phase of K$_2$picene,
similar to electron or hole doping, the superconductivity can occur. The K$_3$-\uppercase\expandafter{\romannumeral1} phase is
a metal with relatively high DOS value of 10.4 states/eV per unit cell at Fermi level. The great difference of electronic
structure between K$_2$-$BD$ and K$_3$-\uppercase\expandafter{\romannumeral1} can lead to different superconducting phases,
and also lead to different pressure effect on the superconducting T$_c$ \cite{Kambe2012}. According to Bardeen-Cooper-Schrieffer
theory of electron-phonon induced superconductivity, the high DOS value of 10.4 states/eV  suggest that
K$_3$-\uppercase\expandafter{\romannumeral1} phase is related to the superconducting phase of T$_c$ = 18 K.

Finally, we analyze the structural feature of K$_3$-\uppercase\expandafter{\romannumeral1} phase. The K atoms located at the
$A,C,E$ positions, which is the most uniform distribution for K$_3$picene.
The length of a hole accommodating K atoms is about 13.76 \AA, equal to the molecule length. It assure that two neighbor K
atoms have an appropriate distance of 4.18$\sim$4.30 \AA, which is close to the distance of 4.54 \AA~ in the bulk metal of
potassium. This uniform distribution can also be found out by the fraction coordinations ($z_1$=0.2074, $z_2$=0.5183 $z_3$=0.8222)
of three K atoms listed in Table. \ref{K2ABC-latt}.
Another feature is that the picene molecule keep in a plane perfectly for K$_3$-\uppercase\expandafter{\romannumeral1} phase
due to K atoms uniform distributions. The superconductivity is correlated intimately to the conjugated $\pi$ orbitals in K-doped
phenanthrene and picene, similar to K or Ca intercalated graphite. The uniform distribution of K atoms helps charge to be
transferred to molecules and the planar molecule with the least distortion helps delocalize $\pi$ electrons throughout
the whole molecule. Therefore, the K$_3$-\uppercase\expandafter{\romannumeral1} phase is favorable for higher SC T$_c$.

In conclusion, we have performed first principle calculations and XRD simulation for K-doped picene.
Our results suggest that K$_2$picene with K$_2$-$BD$ phase is the parent compound of T$_c$ =7 K superconducting phase,
while the K$_3$-\uppercase\expandafter{\romannumeral1} of K$_3$picene corresponds to the superconducting phase with T$_c$ of 18 K. K$_2$-$BD$ phase is semiconducting with 0.1 eV energy gap and K$_3$-\uppercase\expandafter{\romannumeral1} phase is a good metal.
The great difference in electronic properties leads to different T$_c$ and pressure effect on T$_c$ in two superconducting phases.
The existence of KOH in K-doped picene sample can
explain the experimental XRD peaks.
Meanwhile, the generation of KOH is an important factor to
decrease the sample quality.
In a word, we determine the crystal structures of two superconducting phases in K-doped picene, and provide an important basis to further theoretical and experimental studies.

Acknowledgments: We acknowledge Cai-Zhuang Wang for interesting suggestion and fruitful discussion on  XRD simulation.
This work was supported
by MOST 2011CB922200, the Natural Science Foundation of China under Grants
Nos. 91221103, 11174072 and U1204108.


\begin{references}
\bibitem{Mitsuhashi2010} R. Mitsuhashi {\it et al},
 Nature {\bf 464}, 76 (2010).

\bibitem{Wang2011} X. F. Wang {\it et al},
 Nat. Commun. {\bf 2}, 507
(2011).

\bibitem{Xue2012} M. Xue {\it et al},
Sci. Rep. {\bf 2}, 389 (2012).

\bibitem{Wang2011a} X. F. Wang {\it et al},
Phys. Rev. B {\bf 84}, 214523
(2011).

\bibitem{Wang2012} X. F. Wang {\it et al},
J. Phys. Condens.
matter {\bf 24}, 345701 (2012).

\bibitem{Kambe2012} T. Kambe {\it et al},
 Phys. Rev.
B {\bf 86}, 214507 (2012).

\bibitem{Teranishi2013} K. Teranishi {\it et al},
Phys. Rev. B {\bf 87}, 060505 (2013).

\bibitem{Ruff2013} A. Ruff {\it et al},
Phys. Rev. Lett. {\bf 110}, 216403
(2013).

\bibitem{Caputo2012} M. Caputo {\it et al},
J. Phys. Chem. C {\bf 116}, 19902
(2012).


\bibitem{Kosugi2011} T. Kosugi, T. Miyake, S. Ishibashi, R. Arita, and H. Aoki,
Phys. Rev. B {\bf 84}, 214506 (2011).

\bibitem{DeAndres2011a} P. L. de Andres, a. Guijarro, and J. A. Verg\'{e}s, Phys.
Rev. B {\bf 83}, 245113 (2011).

\bibitem{DeAndres2011} P. L. de Andres, a. Guijarro, and J. A. Verg\'{e}s, Phys.
Rev. B {\bf 84}, 144501 (2011).

\bibitem{PhysRevB.88.115106} S. S. Naghavi, M. Fabrizio, T. Qin, and E. Tosatti, Phys.
Rev. B {\bf 88}, 115106 (2013).

\bibitem{Yan2013} X.-W. Yan, Z. Huang, and H.-Q. Lin, J. Chem. Phys.
{\bf 139}, 204709 (2013).

\bibitem{Kim2011} M. Kim, B. I. Min, G. Lee, H. J. Kwon, Y. M. Rhee, and
J. H. Shim, Phys. Rev. B {\bf 83}, 214510 (2011).

\bibitem{Giovannetti2011} G. Giovannetti and M. Capone, Phys. Rev. B 83, 134508
(2011).

\bibitem{Kim2013} M. Kim, H. C. Choi, J. H. Shim, and B. I. Min, New J.
Phys. {\bf 15}, 113030 (2013).

\bibitem{Huang2012} Z. Huang, C. Zhang, and H.-Q. Lin, Sci. Rep. {\bf 2}, 922
(2012).

\bibitem{Verges2012} J. A. Verg\'{e}s {\it et al},
 Phys. Rev. B {\bf 85}, 165102
(2012).

\bibitem{Kato2011} T. Kato, T. Kambe, and Y. Kubozono, Phys. Rev. Lett.
{\bf 107}, 077001 (2011).

\bibitem{PhysRevB.47.558} G. Kresse and J. Hafner, Phys. Rev. B {\bf 47}, 558 (1993).

\bibitem{PhysRevB.54.11169} G. Kresse and J. Furthm\"{u}ller, Phys. Rev. B {\bf 54}, 11169
(1996).

\bibitem{PhysRevB.50.17953} P. E. Bl\"{o}chl, Phys. Rev. B {\bf 50}, 17953 (1994).

\bibitem{PhysRevLett.77.3865} J. P. Perdew, K. Burke, and M. Ernzerhof, Phys. Rev.
Lett. {\bf 77}, 3865 (1996).


\bibitem{mercury} Mercury, Crystal Structure Visualization and Exploration Program,\\
http://www.ccdc.cam.ac.uk/products/mercury (2014).


\end{references}
\end{document}